\documentstyle[11pt,indentfirst]{article}
\newcommand{\be}{\begin{equation}}
\newcommand{\ee}{\end{equation}}

\begin{document}

\title{Criteria for Continuous-Variable Quantum Teleportation}

\author{Samuel L.\ Braunstein$^{\dagger}$, Christopher A.\ Fuchs, and
H.~J. Kimble\medskip\\ \small Norman Bridge Laboratory of Physics,
12-33, \small California Institute of Technology\\ \small Pasadena,
CA 91125, USA\\ \small $^\dagger$SEECS, University of Wales, Bangor
LL57 1UT, UK\\ \small $^\dagger$Hewlett-Packard Labs, Math Group M48,
Bristol BS34 8QZ, UK}

\date{10 April 1999}

\maketitle

\begin{abstract}
We derive an experimentally testable criterion for the teleportation
of quantum states of continuous variables.  This criterion is
especially relevant to the recent experiment of Furusawa {\em et al.}
[Science {\bf 282}, 706 (1998)] where an input-output fidelity of
$0.58\pm0.02$ was achieved for optical coherent states. Our
derivation demonstrates that fidelities greater than $1/2$ could not
have been achieved through the use of a classical channel alone;
quantum entanglement was a crucial ingredient in the experiment.
\end{abstract}

\section{Introduction}

What is quantum teleportation?  The original protocol of Bennett {\it
et al.}~\cite{Bennett93} specifies the idea with succinct clarity.
The task set before Alice and Bob is to transfer the quantum state of
a system in one player's hands onto a system in the other's. The
agreed upon resources for carrying out this task are some previously
shared quantum entanglement and a channel capable of broadcasting
classical information.  It is not allowed to physically carry the
system from one player to the other, and indeed the two players need
not even know each other's locations. One of the most important
features of the protocol is that it must be able to work even when
the state---though perfectly well known to its supplier, a third
party Victor---is completely {\it unknown\/} to both Alice and Bob.
Because the classical information broadcast over the classical
channel can be minuscule in comparison to the infinite amount of
information required to specify the unknown state, it is fair to say
that the state's transport is a disembodied transport
\cite{BraunsteinWeb}. Teleportation has occurred when an unknown
state $|\psi\rangle$ goes in and the same state $|\psi\rangle$ comes
out.

But that is perfect teleportation.  Recent experimental efforts
\cite{Furusawa98,Boschi98,Bouwmeester97,Nielsen98} show there is huge
interest in demonstrating the phenomenon in the laboratory---a venue
where perfection is {\it unattainable\/} as a matter of principle.
The laboratory brings with it a new host of issues:  if perfect
teleportation is unattainable, when can one say that laboratory
teleportation has been achieved?   What appropriate criteria define
the right to proclaim success in an experimental setting? Searching
through the description above, there are several heuristic breaking
points, each asking for quantitative treatment.  The most important
among these are:
\begin{enumerate}
  \item The states should be unknown to Alice and Bob and supplied
  by an actual third party Victor.
  \item Entanglement should be a verifiably used resource, with
  the possibility of physical transportation of the unknown states
  blocked at the outset.   There should be a sense in which the output
  is ``close'' to the input---close enough that it could not have
  been made from information sent through a classical channel alone.
  \item Each and every trial, as defined by Victor's supplying a
  state, should achieve an output sufficiently close to the
  input.  When this situation pertains, the teleportation is called
  {\it unconditional}.  (If that is impractical, {\it conditional\/}
  teleportation---where Alice and Bob are the arbiters of success---may
  still be of interest; but then, at the end of all conditioning, there
  must be a state at the output sufficiently close to the unknown
  input.)
\end{enumerate}
To date only the Furusawa {\em et al.}~experiment \cite{Furusawa98}
has achieved unconditional experimental teleportation as defined by
these three criteria.  The Boschi {\em et al.}~experiment
\cite{Boschi98} fails to meet Criteria (1) and (2) because their
Victor must hand off a (macroscopic) state-preparing device to Alice
instead of an unknown state and because of a variety of low system
efficiencies \cite{Furusawa98}.  The Bouwmeester {\em et
al.}~experiment \cite{Bouwmeester97} fails to meet Criteria (2) and
(3) because their output states---just before they are destroyed by
an extra ``verification'' step---can be produced via communication
through a classical channel alone \cite{Braunstein98a}. In a similar
vein, the Nielsen {\em et al.}~experiment \cite{Nielsen98} fails to
meet these criteria because there is no quantum entanglement shared
between Alice and Bob at any stage of the process
\cite{Braunstein98b,Schack98}.

But the story cannot stop there.  Beside striving for simply better
input-output fidelities or higher efficiencies, there are still
further relevant experimental hurdles to be drawn from
Ref.~\cite{Bennett93}:
\begin{enumerate}
\setcounter{enumi}{3}
  \item The number of bits broadcast over the classical channel
  should be ``minuscule'' in comparison to the information required to
  specify the ``unknown'' states in the class from which the
  demonstration actually draws.
  \item The teleportation quality should be good enough to
  transfer quantum entanglement itself instead of a small subset of
  ``unknown'' quantum states.
  \item The sender and receiver should not have to know each other's
  locations to carry the process through to completion.
\end{enumerate}
And there are likely still more criteria that would seem reasonable
to one or another reader of the original protocol (depending perhaps
upon the particular application called upon). The point is, these two
lists together make it clear that the experimental demonstration of
quantum teleportation cannot be a cut and dried affair.  On the road
toward ideal teleportation, there are significant milestones to be
met and passed.  Important steps have been taken, but the end of the
road is still far from sight.

The work of the theorist in this effort is, among other things, to
help turn the heuristic criteria above into pristine theoretical
protocols within the context of actual experiments.  To this end, we
focus on Criterion 2 in the context of the Furusawa {\em et al.}\
experiment \cite{Furusawa98} where the quantum states of a set of
continuous variables are teleported (as proposed in
Refs.~\cite{Vaidman94,Braunstein98c}). The question is, by what means
can one verify that Alice and Bob---assumed to be at fixed
positions---actually use some quantum entanglement in their purported
teleportation?  How can it be known that they did not use the
resource of a classical channel alone for the quantum state's
transport?  What milestone must be met in order to see this?
Answering these questions fulfills a result already advertised in
Ref.~\cite{Furusawa98} and reported in the abstract of the present
paper.

Our line of attack is to elaborate on an idea first suggested in
Ref.~\cite{Boschi98}.  A cheating Alice and Bob who attempt to make
do with a classical channel alone, must gather information about the
unknown quantum state if they are to have any hope of hiding their
cheat.  But then the limitations of quantum mechanics strike in a
useful way.  As long as the allowed set of inputs contains some
nonorthogonal states, there is no measurement procedure that can
reveal the state's identity with complete reliability.  Any attempt
to reconstruct the unknown quantum state will be flawed necessarily:
information gathering about the identity of a state in a
nonorthogonal set disturbs the state in the process
\cite{Fuchs96,Fuchs98}.  The issue is only to quantify how much
disturbance must take place and to implement the actual comparison
between input and output in an objective, operationally significant
way. If the experimental match (or ``fidelity'') between the input
and output exceeds the bound set by a classical channel, then some
entanglement had to have been used in the teleportation process.

The remainder of the paper is structured as follows.  In the
following section, we discuss the motivation behind choosing the
given measure of fidelity that we do.  We stress in particular the
need for a break with traditional quantum optical measures of signal
transmission, such as signal-to-noise ratio, etc., used in the area
of quantum nondemolition (QND) research \cite{Ralph99,Ralph98}. In
Section 3, we derive the optimal fidelity that can be achieved by a
cheating Alice and Bob whose teleportation measurements are based on
optical heterodyning as in the experiment of Furusawa {\em et
al.}~\cite{Furusawa98}.  This confirms that a fidelity of $1/2$ or
greater is sufficient to assure the satisfaction of Criterion 2 in
that experiment.  We close in Section 4 with a few remarks about some
open problems and future directions.

\section{Why Fidelity?}

Ideal teleportation occurs when an unknown state $|\psi\rangle$ goes
into Alice's possession and the same state $|\psi\rangle$ emerges in
Bob's.  What can this really mean?  A quantum {\it state\/} is not an
objective state of affairs existing completely independently of what
one knows.  Instead it captures the best information available about
how a quantum system will react in this or that experimental
situation \cite{Zeilinger99,Peres93}.\footnote{On this bit of
foundational theory, it seems most experimentalists can agree.  See
in particular page S291 of Zeilinger Ref.~\cite{Zeilinger99} where it
is stated that: ``The quantum state is exactly that representation of
our knowledge of the complete situation which enables the maximal set
of (probabilistic) predictions for any possible future observation. \
\ldots\ \ If we accept that the quantum state is no more than a
representation of the information we have, then the spontaneous
change of the state upon observation, the so-called collapse or
reduction of the wave packet, is just a very natural consequence of
the fact that, upon observation, our information changes and
therefore we have to change our representation of the information,
that is, the quantum state.  From that position, the so-called
measurement problem is not a problem but a consequence of the more
fundamental role information plays in quantum physics as compared to
classical physics.''} This forces one to think carefully about what
it is that is transported in the quantum teleportation process. The
only option is that the teleported $|\psi\rangle$ must {\it always\/}
ultimately refer to someone lurking in the background---a third party
we label Victor, the keeper of knowledge about the system's
preparation. The task of teleportation is to transfer what he can say
about the system he placed in Alice's possession onto a system in
Bob's possession:  it is ``information'' in its purest form that is
teleported, nothing more.

The resources specified for carrying out this task are the previously
shared entanglement between Alice and Bob and a classical channel
with which they communicate.  Alice performs a measurement of a
specified character and communicates her result to Bob.  Bob then
performs a unitary operation on his system based upon that
information.  When Alice and Bob declare that the process is
complete, Victor should know with assurance that whatever his
description of the original system was---his $|\psi\rangle$---it now
holds for the system in Bob's possession.  Knowing with assurance
means that there really is a system that Victor will describe with
$|\psi\rangle$, not that there {\it was\/} a system that he {\it
would have\/} described with $|\psi\rangle$ just before Alice and Bob
declared completion (i.e., as a retrodiction based upon their
pronouncement) \cite{Braunstein98a}.

In any real-world implementation of teleportation, a state
$|\psi_{\rm in}\rangle$ enters Alice and Bob's dominion and a
different state (possibly a mixed-state density operator) $\hat
\rho_{\rm out}$ comes out. As before, one must always keep in mind
that these states refer to what Victor can say about the given system
(see footnote 2).  The question that must be addressed is when
$|\psi_{\rm in}\rangle$ and $\hat\rho_{\rm out}$ are similar enough
to each other that Criterion 2 must have been fulfilled.

We choose to gauge the similarity between $|\psi_{\rm in}\rangle$ and
$\hat\rho_{\rm out}$ by the ``fidelity'' between the two states. This
is defined in the following way\footnote{In order to form this
quantity, we must of course assume a canonical mapping or
identification between the input and output Hilbert spaces.  Any
unitary offset between input and output should be considered a
systematic error, and ultimately taken into account by readjusting
the canonical mapping.  See Ref.~\cite{Ralph99,Ralph98} for a
misunderstanding of this point.  The authors there state, ``\ldots\
fidelity does not necessarily recognize the similarity of states
which differ only by reversible transformations. \ldots\ [This
suggests] that additional measures are required \ldots\ based
specifically on the similarity of measurement results obtained from
the input and output of the teleporter, rather than the inferred
similarity of the input and output states.''  As shown presently, the
fidelity measure we propose does precisely that for all possible
measurements, not just the few that have become the focus of
present-day QND research.}:
\begin{equation}
F\Big(|\psi_{\rm in}\rangle,\hat\rho_{\rm
out}\Big)\equiv\langle\psi_{\rm in}|\hat\rho_{\rm out}|\psi_{\rm
in}\rangle\;. \label{CoffeeBean}
\end{equation}
This measure has the nice property that it equals 1 if and only if
$\hat\rho_{\rm out}=|\psi_{\rm in}\rangle\langle\psi_{\rm in}|$.
Moreover it equals 0 if and only if the input and output states can
be distinguished with certainty by {\it some\/} quantum measurement.
The thing that is really important about this particular measure of
similarity is hinted at by these last two properties.  It captures in
a simple and convenient package the extent to which {\it all
possible\/} measurement statistics produceable by the output state
match the corresponding statistics produceable by the input state.

To see what this means, take {\it any\/} observable (generally a
positive operator-valued measure or POVM \cite{Peres93}) $\{\hat
E_\alpha\}$ with measurement outcomes $\alpha$. If that observable
were performed on the input system, it would give a probability
density for the outcomes $\alpha$ given by
\begin{equation}
P_{\rm in}(\alpha)=\langle\psi_{\rm in}|\hat E_\alpha |\psi_{\rm
in}\rangle\;.
\end{equation}
On the other hand, if the same observable were performed on the
output system, it would give instead a probability density
\begin{equation}
P_{\rm out}(\alpha)={\rm tr}(\hat\rho_{\rm out} \hat E_\alpha)\;.
\end{equation}
A natural way to gauge the similarity of these two probability
densities is by their overlap:
\begin{equation}
{\rm overlap}=\int \sqrt{P_{\rm in}(\alpha)P_{\rm out}(\alpha)} \,
d\alpha\;.
\label{TeaTime}
\end{equation}

It turns out that regardless of which observable is being considered
\cite{Fuchs95,Barnum96},
\begin{equation}
{\rm overlap}^2\ge \langle\psi_{\rm in}|\hat\rho_{\rm out}|\psi_{\rm
in}\rangle\;.
\end{equation}
Moreover there exists an observable that gives precise equality in
this expression \cite{Fuchs95,Barnum96}.  In this sense, the fidelity
captures an operationally defined fact about all possible
measurements on the states in question.

Let us take a moment to stress the importance of a criterion such as
this.  It is not sufficient to attempt to quantify the similarity of
the states with respect to a few observables.  Quantum teleportation
is a much more serious task than classical communication.  Indeed it
is a much more serious task than the simplest forms of quantum
communication, as in quantum key distribution.  In the former case,
one is usually concerned with replicating the statistics of only one
observable across a transmission line.  In the latter case, one is
concerned with reproducing the statistics of a small number of {\it
fixed\/} noncommuting observables (the specific ones required of the
protocol) for a small number of {\it fixed\/} quantum states (the
specific ones required of the protocol).  A full quantum state is so
much more than the quantum measurements in these cases would reveal:
it is a catalog for the outcome statistics of an infinite number of
observables. Good quality teleportation must take that into account.

A concrete example can be drawn from the traditional concerns of
quantum nondemolition measurement (QND) research.  There a typical
problem is how well a communication channel replicates the statistics
of one of two quadratures of a given electromagnetic field mode
\cite{Ralph99,Ralph98}, and most often then only for assumed Gaussian
statistics.  Thinking that quantum teleportation is a simple
generalization of the preservation of signal-to-noise ratio, burdened
only in checking that both quadratures are transmitted faithfully, is
to miss much of the point of teleportation. Specifying the statistics
of two noncommuting observables only goes an infinitesimal way toward
specifying the full quantum state when the Hilbert space is an
infinite dimensional one \cite{Reichenbach44,Vogt78}.

This situation is made acute by noticing that two state vectors can
be almost completely orthogonal---and therefore almost as different
as they can possibly be---while still giving rise to the same $x$
statistics {\it and\/} the same $p$ statistics.  To see an easy
example of this, consider the two state vectors $|\psi_+\rangle$ and
$|\psi_-\rangle$ whose representations in $x$-space are
\begin{equation}
\psi_\pm(x)= \left(\frac{2a}{\pi}\right)^{\! 1/4}\exp\!\left((-a\pm i
b)x^2\right), \label{Herzog}
\end{equation}
for $a,b\ge0$.  In $k$-space representation, these state vectors look
like
\begin{equation}
\tilde\psi_\pm(k)=\left(\frac{a}{2\pi}\right)^{\!
1/4}\sqrt{\frac{a\pm ib}{a^2+b^2}}\, \exp\!\left(\frac{-a\mp
ib}{4(a^2+b^2)}k^2\right).
\end{equation}
Clearly neither $x$ measurements nor $p$ measurements can distinguish
these two states.  For, with respect to both representations, both
wave functions differ only by a local phase function.  However, if we
look at the overlap between the two states we find:
\begin{equation}
\langle\psi_-|\psi_+\rangle=\sqrt{\frac{a(a+ib)}{a^2+b^2}}\;.
\end{equation}
Taking $b\rightarrow\infty$, we can make these two states just as
orthogonal as we please.

Suppose now that $|\psi_+\rangle$ were Victor's input into the
teleportation process, and---by whatever means---$|\psi_-\rangle$
turned out to be the output.  By a criterion that only gauged the
faithfulness of the transmissions of $x$ and $p$ \cite{Ralph99}, this
would be perfect teleportation.  But it certainly isn't so!

Thus the justification of the fidelity measure in
Eq.~(\ref{CoffeeBean}) as a measure of teleportation quality should
be abundantly clear. But this is only the first step in finding a way
to test Criterion 2.  For this, we must invent a quantity that
incorporates information about the teleportation quality of many
possible quantum states.  The reason for this is evident: in general
it is possible to achieve a nonzero fidelity between input and output
even when a cheating Alice and Bob use no entanglement whatsoever in
their purported teleportation.  This can come about whenever Alice
and Bob can make use of some prior knowledge about Victor's actions.

As an example, consider the case where Alice and Bob are privy to the
fact that Victor wishes only to teleport states drawn from a given
{\it orthogonal\/} set.  At any shot, they know they will be given
one of these states, just not which one.  Then, clearly, they need
use no entanglement to ``transmit'' the quantum states from one
position to the other.  A cheating Alice need only perform a
measurement $\cal O$ whose eigenstates coincide with the orthogonal
set and send the outcome she obtains to Bob. Bob can use that
information to resynthesize the appropriate state at his end. No
entanglement has been used, and yet with respect to these states
perfect teleportation has occurred.

This example helps define the issue much more sharply.  The issue
turns on having a general statement of what it means to say that
Alice and Bob are given an {\it unknown\/} quantum state?   In the
most general setting it means that Alice and Bob know that Victor
draws his states $|\psi_{\rm in}\rangle$ from a fixed set $\cal S$;
they just know not which one he will draw at any shot.  This lack of
knowledge is taken into account by a probability ascription
$P(|\psi_{\rm in}\rangle)$.  That is:
\begin{quote}
\it
All useful criteria for the achievement of teleportation must be
anchored in whatever $\cal S$ and $P(|\psi_{\rm in}\rangle)$ are
given.  A criterion is senseless if the states to which it is to be
applied are not mentioned explicitly.
\end{quote}
This makes it sensible to consider the average fidelity between input
and output
\begin{equation}
F_{\rm av} = \int_{\cal S}P(|\psi_{\rm in}\rangle)\, F\Big(|\psi_{\rm
in}\rangle,\hat\rho_{\rm out}\Big)\, d|\psi_{\rm in}\rangle\;,
\end{equation}
as a benchmark capable of eliciting the degree to which Criterion 2
is satisfied. If $\cal S$ consists of orthogonal states, then no
criterion whatsoever (short of watching Alice and Bob's every move)
will ever be able to draw a distinction between true teleportation
and the sole use of the classical side channel.  Things only become
interesting when the set $\cal S$ consists of two or more {\it
nonorthogonal\/} quantum states \cite{Boschi98}:  for only then will
$F_{\rm av}=1$ {\it never\/} be achievable by a cheating Alice and
Bob.

By making the set $\cal S$ more and more complicated, we can define
ever more stringent tests connected to Criterion 2.  For instance,
consider the simplest nontrivial case:  take ${\cal S}={\cal S}_0=
\{|\psi_0\rangle,|\psi_1\rangle\}$, a set of just two nonorthogonal
states (with a real inner product $x=\cos\theta$). Suppose the two
states occur with equal probability.  Then it can be shown
\cite{Fuchs96} that the best thing for a cheating Alice and Bob to do
is this.  Alice measures an operator whose orthogonal eigenvectors
symmetrically bestride $|\psi_0\rangle$ and $|\psi_1\rangle$.  Using
that information, Bob synthesizes one of two states
$|\tilde\psi_0\rangle$ and $|\tilde\psi_1\rangle$ each lying in the
same plane as the original two states, but each tweaked slightly
toward the other by an angle \cite{Fuchs96b}
\begin{equation}
\phi=\frac{1}{2}\arctan\!\left[
\left({1+\sin\theta\over1-\sin\theta}+\cos2\theta\right)^{\!\!-1}\!\sin2\theta
\right]\;.
\end{equation}
This (optimal) strategy gives a fidelity
\begin{equation}
F_{\rm av}=\frac{1}{2}\!\left(1+\sqrt{1-x^2+x^4}\right)\,.
\label{herpolhode}
\end{equation}
Even in the worst case (when $x=1/\sqrt{2}$), this fidelity is always
relatively high---it is always above 0.933 \cite{Fuchs99}.

This shows that choosing ${\cal S}_0$ to check for the fulfillment of
Criterion 2 is a very weak test.  For an example of the opposite
extreme, consider the case where $\cal S$ consists of every
normalized vector in a Hilbert space of dimension $d$ and assume that
$\cal S$ is equipped with the uniform probability distribution (i.e.,
the unique distribution that is invariant with respect to all unitary
operations). Then it turns out that the maximum value $F_{\rm av}$
can take is \cite{Barnum98}
\begin{equation}
F_{\rm av}=\frac{2}{d+1}\,.
\end{equation}
For the case of a single qubit, i.e., $d=2$, Alice and Bob would {\it
only\/} have to achieve a fidelity of $2/3$ before they could claim
that they verifiably used some entanglement for their claimed
teleportation. But, again, this is only if Victor can be sure that
Alice and Bob know absolutely nothing about which state he inputs
other than the dimension of the Hilbert space it lives in.

This last example finally prepares us to build a useful criterion for
the verification of continuous quantum-variable teleportation in the
experiment of Furusawa {\it et al.}~\cite{Furusawa98}.  For a
completely unknown quantum state in that experiment would correspond
to taking the limit $d\rightarrow\infty$ above. If Victor can be sure
that Alice and Bob know nothing whatsoever about the quantum states
he intends to teleport, then on average the best fidelity they can
achieve in cheating is strictly zero!  In this case, seeing any
nonzero fidelity whatsoever in the laboratory would signify that
unconditional quantum teleportation had been achieved.

But making such a drastic assumption for the confirmation set $\cal
S$ would be going too far.  This would be the case if for no other
reason because any present-day Victor lacks the experimental ability
to make good his threat.  Any Alice and Bob that had wanted to cheat
in the Furusawa {\it et al}.\ experiment would know that the Victor
using their services is technically restricted by the fact that only
a handful of manifestly quantum or nonclassical states have ever been
generated in quantum optics laboratories \cite{Mandel95}. By far the
most realistic and readily available laboratory source available to
Victor is one that creates optical coherent states of a single field
mode for his test of teleportation. Therefore in all that follows we
will explicitly make the assumption that $\cal S$ contains the
coherent states $|\alpha\rangle$ with a Gaussian distribution
centered over the vacuum state describing the probability density on
that set.  As we shall see presently, it turns out that in the limit
that the variance of the Gaussian distribution approaches
infinity---i.e., the distribution of states becomes ever more
uniform---the upper bound for the average fidelity achievable by a
cheating Alice and Bob using optical heterodyne measurements is
\begin{equation}
F_{\rm av}=\frac{1}{2}\;.
\label{polhode}
\end{equation}
Any average fidelity that exceeds this bound must have come about
through the use of some entanglement.

\section{Optimal Heterodyne Cheating}

We now verify Eq.~(\ref{polhode}) within the context of the Furusawa
{\it et al}.\ experiment.  There, the object is to teleport an
arbitrary coherent state of a finite bandwidth electromagnetic field.
(The extension of the single mode theory of Ref.~\cite{Braunstein98c}
to the multimode case is given in Ref.~\cite{vanLoock99}.)  We focus
for simplicity on the single mode case.  The quantum resource used
for the process is one that entangles the number states $|n\rangle$
of two modes of the field. Explicitly the entangled state is given by
\cite{vanEnk99}
\begin{equation}
|E\rangle_{\rm\scriptscriptstyle AB}=\frac{1}{\cosh
r}\sum_{n=0}^\infty (\tanh r)^n |n\rangle_{\rm\scriptscriptstyle A}
|n\rangle_{\rm\scriptscriptstyle B}\;,
\end{equation}
where $r$ measures the amount of squeezing required to produce the
entangled state.

In order to verify that entanglement was actually used in the
experiment, as discussed in the previous section, we shall assume
that the test set $\cal S$ is the full set of coherent states
$|\beta\rangle$,
\begin{equation}
|\beta\rangle=\exp(-|\beta|^2/2)
\sum_0^\infty\frac{\beta^n}{\sqrt{n!}}|n\rangle\;,
\end{equation}
where the complex parameter $\beta$ is distributed according to a
Gaussian distribution,
\begin{equation}
p(\beta)=\frac{\lambda}{\pi}e^{-\lambda|\beta|^2}\;.
\end{equation}
Ultimately, of course, we would like to consider the case where Alice
and Bob are completely ignorant of which coherent state is drawn.
This is described by taking the limit $\lambda\rightarrow0$ in what
follows.

It is well known that the measurement optimal for estimating the
unknown parameter $\beta$ when it is distributed according to a
Gaussian distribution \cite{Yuen73} is the POVM $\{\hat E_\alpha\}$
constructed from the coherent state projectors according to
\begin{equation}
\hat E_\alpha=\frac{1}{\pi}|\alpha\rangle\langle\alpha|\;,
\end{equation}
first suggested by Arthurs and Kelly \cite{Arthurs65}.  This
measurement is equivalent to optical heterodyning \cite{Personick71}.
These points make this measurement immediately attractive for the
present considerations.  On the one hand, maximizing the average
fidelity (as is being considered here) is almost identical in spirit
to the state-estimation problem of Ref.~\cite{Yuen73}.  On the other,
in the Furusawa {\it et al.} experiment a cheating Alice who uses no
entanglement actually performs precisely this measurement.

We therefore consider an Alice who performs the measurement $\{\hat
E_\alpha\}$ and forwards on the outcome---i.e., the complex number
$\alpha$---to Bob.\footnote{We caution however that the present
considerations do not {\it prove\/} the optimality of heterodyne
measurement for an arbitrarily adversarial Alice and Bob---they
simply make it fairly plausible.  Complete optimization requires the
consideration of all POVMs that Alice can conceivably perform along
with explicit consideration of the structure of the fidelity function
considered here, not simply the variance of an estimator as in the
state-estimation problem.  More on this issue can be found in
Ref.~\cite{Fuchs99z}.} The only thing Bob can do with this
information is generate a new quantum state according to some rule,
$\alpha \rightarrow |f_\alpha\rangle$. Let us make no {\it a
priori\/} restrictions on the states $|f_\alpha\rangle$.  The task is
first to find the maximum average fidelity $F_{\rm max}(\lambda)$ Bob
can achieve for a given $\lambda$.

For a given strategy $\alpha \rightarrow |f_\alpha\rangle$, the
achievable average fidelity is
\begin{eqnarray}
F(\lambda) &=& \int p(\beta)\!\left(\int p(\alpha|\beta)\, |\langle
f_\alpha|\beta\rangle|^2 d^2\alpha\right)\!d^2\beta
\\
&=& \int
p(\beta)\!\left(\int\frac{1}{\pi}|\langle\alpha|\beta\rangle|^2
|\langle f_\alpha|\beta\rangle|^2 d^2\alpha\right)\!d^2\beta
\\
&=&
\frac{\lambda}{\pi^2}\int\int e^{-\lambda|\beta|^2}
e^{-|\alpha-\beta|^2}|\langle f_\alpha|\beta\rangle|^2\, d^2\beta\,
d^2\alpha
\\
&=& \frac{\lambda}{\pi^2}\int e^{-|\alpha|^2}\langle
f_\alpha|\!\left( \int\exp\!\Big(-(1+\lambda)|\beta|^2+2{\rm
Re}\,\alpha^*\beta \Big)
|\beta\rangle\langle\beta|\,d^2\beta\right)|f_\alpha\rangle\,
d^2\alpha\;.
\label{ShutUp--GoAway}
\end{eqnarray}

Notice that the operator enclosed within the brackets in
Eq.~(\ref{ShutUp--GoAway}), i.e.,
\begin{equation}
\hat{\cal O}_\alpha=\int\exp\!\Big(-(1+\lambda)|\beta|^2+ 2{\rm
Re}\,\alpha^*\beta\Big)\, |\beta\rangle\langle\beta|\, d^2\beta\;,
\end{equation}
is a positive semi-definite Hermitian operator that depends only on
the real parameter $\lambda$ and the complex parameter $\alpha$. It
follows that
\begin{equation}
\langle f_\alpha|\hat{\cal O}_\alpha| f_\alpha\rangle\le
\mu_1(\hat{\cal O}_\alpha)\;,
\label{Macaroni}
\end{equation}
where $\mu_1(\hat X)$ denotes the largest eigenvalue of the operator
$\hat X$.

With this, Bob's best strategy is apparent.  For each $\alpha$, he
simply adjusts the state $| f_\alpha\rangle$ to be the eigenvector of
$\hat{\cal O}_\alpha$ with the largest eigenvalue.  Then equality is
achieved in Eq.~(\ref{Macaroni}), and it is just a question of being
able to perform the integral in Eq.~(\ref{ShutUp--GoAway}).

The first step in carrying this out is to find the eigenvector and
eigenvalue achieving equality in Eq.~(\ref{Macaroni}). This is most
easily evaluated by unitarily transforming $\hat{\cal O}_\alpha$ into
something that is diagonal in the number basis, picking off the
largest eigenvalue, and transforming back to get the optimal $|
f_\alpha\rangle$.  (Recall that eigenvalues are invariant under
unitary transformations.)

The upshot of this procedure is best illustrated by working backward
toward the answer. Consider the positive operator
\begin{equation}
\hat P=\int e^{-(1+\lambda)|\beta|^2} |\beta\rangle\langle\beta|\,
d^2\beta\;.
\end{equation}
Expanding this operator in the number basis, we find
\begin{equation}
\hat P=\pi\sum_{n=0}^\infty (2+\lambda)^{-(n+1)}
|n\rangle\langle n|\;.
\end{equation}
So clearly,
\begin{equation}
\mu_1(\hat P)=\frac{\pi}{2+\lambda}\;.
\end{equation}

Now consider the displaced operator
\begin{equation}
\hat Q_\alpha=\hat D\!\left(\frac{\alpha}{1+\lambda}\right)\hat P
\,\hat D^\dagger\!\left(\frac{\alpha}{1+\lambda}\right)\;,
\end{equation}
where $\hat D(\nu)$ is the standard displacement operator
\cite{Nussenzveig73}. Working this out in the coherent-state basis,
one finds
\begin{eqnarray}
\hat Q_\alpha
&=&
\int e^{-(1+\lambda)|\beta|^2}\left|\beta+
\frac{\alpha}{1+\lambda}\right\rangle
\left\langle\beta+
\frac{\alpha}{1+\lambda}\right|d^2\beta
\\
&=& \int
\exp\!\left(-(1+\lambda)\left|\gamma-\frac{\alpha}{1+\lambda}
\right|^2\,\right)|\gamma\rangle\langle\gamma|\,d^2\gamma
\\
&=& \exp\!\left(\frac{-|\alpha|^2}{1+\lambda}\right)
\int\exp\!\Big(-(1+\lambda)|\gamma|^2+ 2{\rm Re}\,\alpha^*\gamma\Big)
|\gamma\rangle\langle\gamma|\, d^2\gamma
\\
&=&
\exp\!\left(\frac{-|\alpha|^2}{1+\lambda}\right)
\hat{\cal O}_\alpha\;.
\end{eqnarray}
Using this in the expression for $F(\lambda)$ we find,
\begin{eqnarray}
F(\lambda) &=&
\frac{\lambda}{\pi^2}\int\exp\!\left(-\left(1-\frac{1}{1+\lambda}
\right)|\alpha|^2\right)\langle f_\alpha|\left( \hat
D\!\left(\frac{\alpha}{1+\lambda}\right)\hat P \,\hat
D^\dagger\!\left(\frac{\alpha}{1+\lambda}\right)\right)|f_\alpha
\rangle\,d^2\alpha
\\
&\le&
\frac{1}{\pi}\frac{\lambda}{2+\lambda}\int\exp\!
\left(-\frac{\lambda}{1+\lambda}|\alpha|^2\right)d^2\alpha
\\
&=&
\frac{1+\lambda}{2+\lambda}\;.
\end{eqnarray}
Equality is achieved in this chain by taking
\begin{equation}
|f_\alpha\rangle=
D\!\left(\frac{\alpha}{1+\lambda}\right)|0\rangle=
\left|\frac{\alpha}{1+\lambda}\right\rangle\;.
\label{Cheese}
\end{equation}
Therefore the maximum average fidelity is given by
\begin{equation}
F_{\rm max}(\lambda)=\frac{1+\lambda}{2+\lambda}\;.
\end{equation}
In the limit that $\lambda\rightarrow0$, i.e., when Victor draws his
states from a uniform distribution, we have
\begin{equation}
F_{\rm max}(\lambda)\;\longrightarrow\; \frac{1}{2}\;,
\end{equation}
as advertised in Ref.~\cite{Furusawa98}.

It should be noted that nothing in this argument depended upon the
mean of the Gaussian distribution being $\beta=0$.  Bob would need to
minimally modify his strategy to take into account Gaussians with a
non-vacuum state mean, but the optimal fidelity would remain the
same.

\section{Conclusion}

Where do we stand?  What remains?  Clearly one would like to develop
a toolbox of ever more stringent and significant tests of quantum
teleportation---ones devoted not only to Criterion 2, but to all the
others mentioned in the Introduction as well.  Significant among
these are delineations of the fidelities that must be achieved to
ensure the honest teleportation of nonclassical states of light, such
as squeezed states.  Some work in this direction appears in
Ref.~\cite{Braunstein98c}, but one would like to find something more
in line with the framework presented here.  Luckily, a more general
setting for this problem can be formulated \cite{Fuchs99z} as it will
ultimately be necessary to explore any number of natural verification
sets $\cal S$ and their resilience with respect to arbitrarily
adversarial Alice and Bob teams.

\section{Acknowledgments}
We thank Jason McKeever for suggesting the nice example in
Eq.~(\ref{Herzog}) and thank J.~R. Buck and C.~M. Caves for useful
discussions.  This work was supported by the QUIC Institute funded by
DARPA via the ARO, by the ONR, and by the NSF\@.  SLB was funded in
part by EPSRC grant GR/L91344. CAF acknowledges support of the Lee A.
DuBridge Fellowship.


\begin{thebibliography}{99}
\bibitem{Bennett93}
C.~H. Bennett, G.~Brassard, C.~Cr\'{e}peau, R.~Jozsa, A.~Peres, and
W.~K. Wootters, ``Teleporting an Unknown Quantum State via Dual
Classical and {E}instein-{P}odolsky-{R}osen Channels,'' Phys.\ Rev.\
Lett.\ {\bf 70}, 1895 (1993).

\bibitem{BraunsteinWeb}
S.~L. Braunstein, ``A Fun Talk on Teleportation,'' available on the
World Wide Web at {\tt http://
www.sees.bangor.ac.uk/$\tilde{\;\;}\!$schmuel/tport.html}.

\bibitem{Furusawa98}
A.~Furusawa, J.~L. S{\o}rensen, S.~L. Braunstein, C.~A. Fuchs,
H.~J. Kimble, and E.~S. Polzik, ``Unconditional Quantum
Teleportation,'' Science {\bf 282}, 706 (1998).

\bibitem{Boschi98}
D.~Boschi, S.~Branca, F.~De~Martini, L.~Hardy, and S.~Popescu,
``Experimental Realization of Teleporting an Unknown Pure Quantum
State via Dual Classical and Einstein-Podolsky-Rosen Channels,''
Phys.\ Rev.\ Lett.\ {\bf 80}, 1121--1125 (1998).

\bibitem{Bouwmeester97}
D.~Bouwmeester, J.-W.~Pan, K.~Mattle, M.~Eibl, H.~Weinfurter, and
A.~Zeilinger, ``Experimental Quantum Teleportation,'' Nature {\bf
390}, 575--579 (1997).

\bibitem{Nielsen98}
M.~A. Nielsen, E. Knill, and R.~Laflamme, ``Complete Quantum
Teleportation Using Nuclear Magnetic Resonance,'' Nature {\bf
396}, 52--55 (1998).

\bibitem{Braunstein98a}
S.~L. Braunstein and H.~J. Kimble, ``{\em A Posteriori}
Teleportation,'' Nature {\bf 394}, 840--841 (1998).

\bibitem{Braunstein98b}
S.~L. Braunstein, C.~M. Caves, R.~Jozsa, N.~Linden, S.~Popescu,
and R. Schack, ``Separability of Very Noisy Mixed States and
Implications for NMR Quantum Computing,'' {\tt quant-ph/9811018}.

\bibitem{Schack98}
R.~Schack and C.~M. Caves, ``Classical Model for Bulk-Ensemble NMR
Quantum Computation,'' {\tt quant-ph/9903101}.

\bibitem{Vaidman94}
L.~Vaidman, ``Teleportation of Quantum States,'' Phys.\ Rev.\ A
{\bf 49}, 1473--1476 (1994).

\bibitem{Braunstein98c}
S.~L. Braunstein and H.~J. Kimble, ``Teleportation of Continuous
Quantum Variables,'' Phys.\ Rev.\ Lett.\ {\bf 80}, 869--872
(1998).

\bibitem{Fuchs96}
C.~A. Fuchs and A. Peres, ``Quantum State Disturbance vs.\
Information Gain: Uncertainty Relations for Quantum Information,''
Phys.\ Rev.\ A {\bf 53}, 2038--2045 (1996).

\bibitem{Fuchs98}
C.~A. Fuchs, ``Information Gain vs.\ State Disturbance in Quantum
Theory,'' Fort.\ Phys.\ {\bf 46} 535--565 (1998).

\bibitem{Ralph99}
T.~C. Ralph, P.~K. Lam, and R.~E.~S. Polkinghorne, ``Characterizing
Teleportation in Optics,'' {\tt quant-ph/9903003}.

\bibitem{Ralph98}
T.~C. Ralph and P.~K. Lam, ``Teleportation with Bright Squeezed
Light,'' Phys.\ Rev.\ Lett. {\bf 81}, 5668--5671 (1998).

\bibitem{Zeilinger99}
A.~Zeilinger, ``Experiment and the Foundations of Quantum Physics,''
Rev.\ Mod.\ Phys.\ {\bf 71}, S288--S297 (1999).

\bibitem{Peres93}
A.~Peres, {\sl Quantum Theory: Concepts and Methods}, (Kluwer,
Dordrecht, 1993).

\bibitem{Fuchs95}
C.~A. Fuchs and C.~M. Caves, ``Mathematical Techniques for Quantum
Communication Theory,'' Open Sys.\ and Info.\ Dyn.\ {\bf 3}, 345--356
(1995).

\bibitem{Barnum96}
H.~Barnum, C.~M. Caves, C.~A. Fuchs, R.~Jozsa, and B.~Schumacher,
``Noncommuting Mixed States Cannot Be Broadcast,'' Phys.\ Rev.\
Lett.\ {\bf 76}, 2818--2821 (1996).

\bibitem{Reichenbach44}
H.~Reichenbach, {\sl Philosophic Foundations of Quantum Mechanics},
(U. California Press, Berkeley, 1944), pp.~91--92.

\bibitem{Vogt78}
A.~Vogt, ``Position and Momentum Distributions Do Not Determine the
Quantum Mechanical State,'' in {\sl Mathematical Foundations of
Quantum Theory}, edited by A.~R. Marlow (Academic Press, New York,
1978), pp.~365--372.

\bibitem{Fuchs96b}
C.~A. Fuchs, {\sl Distinguishability and Accessible Information in
Quantum Theory}, Ph.~D. thesis, University of New Mexico,
Albuquerque, NM (1996); see also {\tt quant-ph/9601020}.

\bibitem{Fuchs99}
C.~A. Fuchs, ``Just {\it Two\/} Nonorthogonal Quantum States,'' to
appear in {\sl Quantum Communication, Computing, and Measurement 2},
edited by P.~Kumar, G.~M. D'Ariano, and O.~Hirota (Plenum Press, NY,
1998); see also {\tt quant-ph/9810032}.

\bibitem{Barnum98}
H.~Barnum, {\sl Quantum Information}, Ph.~D. thesis, University of
New Mexico, Albuquerque, NM (1998).

\bibitem{Mandel95}
L.~Mandel and E.~Wolf, {\sl Optical Coherence and Quantum Optics},
(Cambridge U. Press, Cambridge, 1995).

\bibitem{vanLoock99}
P.~van Loock, S.~L. Braunstein, and H.~J. Kimble, ``Broadband
Teleportation,'' to appear in Phs.\ Rev.\ A; see also {\tt
quant-ph/9902030}.

\bibitem{vanEnk99}
S.~J. van Enk, ``On Teleportation with Continuous Variables,'' to
appear (1999).

\bibitem{Yuen73}
H.~P. Yuen and M.~Lax, ``Multiple-Parameter Quantum Estimation and
Measurement of Nonselfadjoint Observables,'' IEEE Trans.\ Inf.\
Theor.\ {\bf IT-19}, 740--750 (1973).

\bibitem{Arthurs65}
E.~Arthurs and J.~L. Kelly, Jr., ``On the Simultaneous Measurement of
a Pair of Conjugate Observables,'' Bell Syst.\ Tech.\ J. {\bf 44},
725--729 (1965).

\bibitem{Personick71}
S.~D. Personick, ``An Image-Band Interpretation of Optical Heterodyne
Noise,'' Bell Syst.\ Tech.\ J. {\bf 50}, 213--216 (1971).

\bibitem{Nussenzveig73}
H.~M. Nussenzveig, {\sl Introduction to Quantum Optics}, (Gordon and
Breach, London, 1973).

\bibitem{Fuchs99z}
C.~A. Fuchs, ``Squeezing Quantum Information through a Classical
Channel,'' in preparation.

\end{thebibliography}
\end{document}